\documentclass[b5paper,twoside]{jpconf}  
\usepackage{graphicx}

\begin{document}

\title[LCGT and gravitational wave detectors]{LCGT and the global network of gravitational wave detectors}  

\author[N.Kanda]{Nobuyuki Kanda$^1$ and LCGT collaboration}

\address{$^1$Department of Physics, Graduate School of Science, Osaka City University, Sugimoto 3-3-138, Sumiyoshi-ku, Osaka 558-8585, Japan}
\ead{kanda@sci.osaka-cu.ac.jp} 

\begin{abstract}
Gravitational wave is a propagation of space-time distortion, which is predicted by Einstein in general relativity. Strong gravitational waves will come from some drastic astronomical objects, e.g. coalescence of neutron star binaries, black holes, supernovae, rotating pulsars and pulsar glitches. Detection of the gravitational waves from these objects will open a new door of \textit{`gravitational wave astronomy'}. Gravitational wave will be a probe to study the physics and astrophysics.

To search these gravitational waves, large-scale laser interferometers will compose a global network of detectors. Advanced LIGO and advanced Virgo are upgrading from currents detectors. One of LIGO detector is considering to move Australia Site. IndIGO or Einstein Telescope are future plans.

LCGT (Large-scale Cryogenic Gravitational wave Telescope) is now constructing in Japan with distinctive characters: cryogenic cooling mirror and underground site. We will present a design and a construction status of LCGT, and brief status of current gravitational wave detectors in the world.

Network of these gravitational wave detectors will start in late 2016 or 2017, and may discover the gravitational waves. For example, these detectors will reach its search range for coalescence of neutron star binary is over 200 Mpc, and several or more events per year will be expected. Since most of gravitational wave events are from high-energy phenomenon of the astronomical objects, these might have counterpart evidences in electromagnetic radiation (visible light, X/gamma ray), neutrino, high energy particles or others. Thus, the mutual follow-up observations will give us more information of these objects. 
\end{abstract}

\section{Introduction : Gravitational Waves}

Gravitational waves are predicted by Einstein's general relativistic theory in 1916. With the perturbation of space-time metric in Einsteins's equation, a wave equation is derived as
\begin{equation}
\left({{\nabla }^{2}\ -\ {\frac{1}{c}}{\frac{{\partial }^{2}\,}{\partial 
\mit \,{t}^{2}}}}\right)\ {h}_{\mu \nu }\ =\ 0
\label{eq:GWEq}
\end{equation}
where, $h_{\mu \nu}$ respects the metric perturbation away from Minkowski metric as
\begin{equation}
{g}_{\mu \nu }\ ={\eta }_{\mu \nu }\ +\ {h}_{\mu \nu }\ .
\label{eq:hDefinition}
\end{equation}
This equation suggest that small distortion of space-time $h_{\mu \nu}$ will be able to propagate as wave, which is called as {\it ``gravitational wave''}. 

The gravitational wave will emitted from a changing of mass quadrupole moment, just like as a non-spherical motion of mass distribution. The lowest order of a gravitational wave is quadrupole radiation. With transverse-traceless gauge, choosing $z$ axis as a propagation, the wave can be represented the linear combination of two polarization :
\begin{equation}
{h}_{+}\ =\ a \left({\matrix{0&0&0&0\cr
0&1&0&0\cr
0&0&-1&0\cr
0&0&0&0\cr}}\right)\ ,\ \ 
{h}_{\rm \times }\ =\ b \left({\matrix{0&0&0&0\cr
0&0&1&0\cr
0&1&0&0\cr
0&0&0&0\cr}}\right)\ .
\label{eq:hCross}
\end{equation}
It is a transverse wave, and it propagate with light speed.
A gravitational wave will interact with masses. When it incident, a tidal force will be induced on free falling masses. The $h_{+}$ component will induce the distortion along $x$- and $y$-axis differentially, and $h_{\times}$ will induce same but rotating 45 degree.

Unit of gravitational wave amplitude is dimensionless metric. It means that space-time distortion relativity.
The gravitational interaction is very weak comparing with other fundamental inter, and also the gravitational wave amplitude is extremely small as
\begin{equation}
h_{\mu\nu} \ =\ \frac{2G}{Rc^4} \ddot{I}_{\mu\nu}
\end{equation}
, where $I_{\mu\nu}$ is a quadrupole moment of mass distribution. For example, two 100 kg masses connected with 1m beam, rotating 1000 cycles per second will emit 2kHz gravitational wave of amplitude only $\sim 10^{-40}$ at a distance of 150km where is wavelength.

\subsection{Gravitational Waves from Astronomical Objects}

Since the gravitational wave is weak, we expect sources consists of huge mass motions - astronomical objects. Also the sources must have large acceleration to make a large derivative of quadrupole motion. Typical sources of gravitational waves are coalescence of compact star (neutron star, black hole) binaries, stellar-core collapse of supernovae.
These are occasionally gravitational wave sources. On the other hand, rotating pulsars or binaries might be a sources of continuous gravitational waves.

Neutron star binary is a most promising gravitational wave sources. PSR1913+16 is a binary pulsar is known as that their Kepler orbit period changing is well consistent with a transfer of energy and momentum from binary system by gravitational wave radiation\cite{TaylorHulse}. There are some observations of such binaries\cite{lrr-2008}. A Binary system will loss the energy emitting gravitational waves, will shrink its orbit, and become rotate faster. Finally, the binary coalesce and emit large gravitational wave. It is called as {\it `the last three minutes'}\cite{TheLastThreeMin}. The amplitude $h$ of gravitational wave neutron star binary from 200 Mpc away will be $10^{-24}\sim 10^{-23}$ in frequency band of 100Hz -- 1kHz, and
its frequency spectral density will be $h(f) \sim 10^{-22} \sim 10^{-23} {\rm [1/\sqrt{Hz}]}$. Gravitational waveforms of compact binary coalescence can be well predicted by post-newtonian \cite{CBCwaveform}. The frequency and amplitude development along the time ate determined by the masses of stars. Thus, once we measure the waveform, it guess the absolute amplitude of gravitational wave from the source. Comparing observed amplitude, we can estimate the luminosity distance of sources only by gravitational wave measurement. In this mean, compact binary coalescence are called as {\it `standard candle (siren)'}.
The rate of coalescence of neutron star binary is estimated as $118^{+174}_{-79}$ events/Myr/Galaxy \cite{Kim_GMR}. Even this expectation contains large statistical error and dependency of some models, it is encouraging us to try to detect the gravitational waves from compact star coalescence.

On the other hand, the weakness, i.e. small coupling constant of gravitational wave make possible that its come from an inside the star. Electromagnetic radiations are mostly emitted at the surface of stars. Neutrino may come from inside as a core of supernovae. Gravitational wave come from mass motion itself inside the stars, and pass through the materials. This fact suggest that gravitational wave will bring many information of deep inside of high-energetic astronomical object. In the case of black hole, only a gravitational wave is radiation from themselves. We expect a new window of {\it `gravitational wave astronomy'} will be opened with current constructing/upgrading gravitational wave detectors.

\begin{figure}[bt]
\begin{center}
\includegraphics[width=80mm,trim=0 10 0 0, clip]{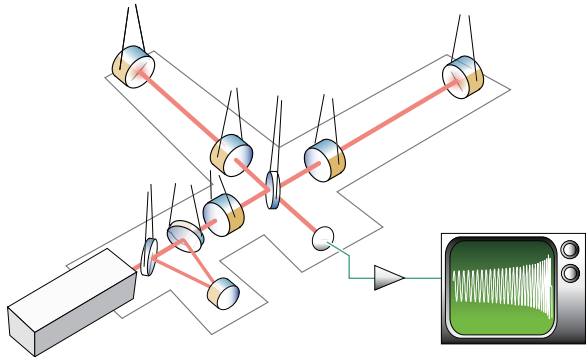}
\end{center}
\vspace{-5mm}
\caption{\label{IFOske}Schematic view of laser interferometric gravitational wave detector}
\end{figure}

\subsection{Detection Principle of Gravitational Waves}

Gravitational wave induce the tidal force between free masses. Laser interferometric gravitational wave detectors are Michelson interferometer that consists of suspended mirrors, beam splitter and laser as light source. Mirrors and beamsplitter are suspended as pendulum that is a mechanical low pass filter which make mirror as a free mass. Laser light is divided to two perpendicular paths,
round trip from mirrors and re-combine at the beam splitter. We can measure the metric amplitude of gravitational wave by small change of intensity of the combined light. To enhance a gravitational wave signal, long base-line of the Michelson beam path and Fabry-Perot cavity are employed. Figure \ref{IFOske} shows the schematic view of the detector.

\section{LCGT and Other Gravitational Wave Detectors} 

\subsection{LCGT}
\begin{figure}[htb]
\begin{center}
\includegraphics[width=80mm]{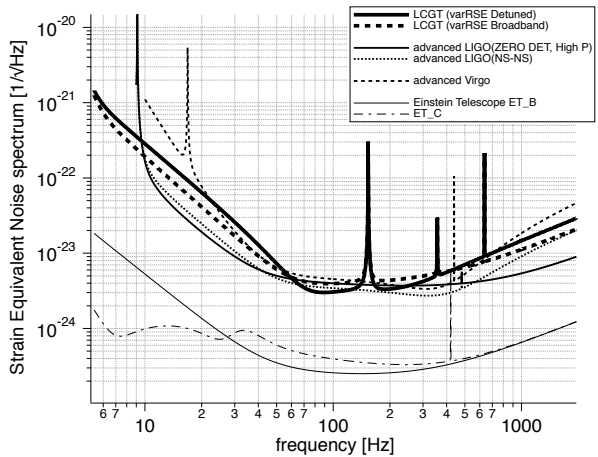}
\end{center}
\vspace{-5mm}
\caption{\label{sens}Target sensitivity limits of current constructing/upgrading/planning gravitational wave detectors. Design values of each detectors are available in references\cite{spe_LCGT}\cite{spe_aLIGO}\cite{spe_aVirgo}\cite{spe_aET}.}
\end{figure}

 LCGT (\textbf{L}large-scale \textbf{C}ryogenic \textbf{G}ravitational wave \textbf{T}elescop ) is a 3 km base-line length laser interferoemeteric detector at the Kamioka-mine, Gifu-prefecture, Japan\cite{LCGT}. As its name shows, LCGT consists of cryogenic mirrors to reduce the thermal motion noise of mirrors. The site of the LCGT is underground for stable environment with low seismic noises magnitude comparing with surface of the ground. 
LCGT is planning to start its observational operation in 2017.
A sensitivity of gravitational wave detector is characterized by an strain equivalent noise spectrum.
LCGT's target sensitivity which just corresponding to the strain equivalent noise level is $h \sim factor \times 10^{-24} {\rm [1/\sqrt{Hz}]}$ around 100 Hz as same to other current upgrading gravitational wave detectors as shown in figure \ref{sens} with advanced LIGO, advanced Virgo and ET\cite{spe_LCGT}\cite{spe_aLIGO}\cite{spe_aVirgo}\cite{spe_aET}.

\subsection{Other Gravitational Wave Detectors in Current}

There are some gravitational wave detector projects. LIGO\cite{LIGO} (Laser Interferometer Gravitational-Wave Observatory) is a US project with two dislocated detector site at west and east side of north American continent. Virgo\cite{VIRGO} is a european (mainly from Italy and France) project at Pisa, Italy. These detectors achieved its `initial' operation already, and now are upgrading as each sensitivity reach as $200 \sim 300$ Mpc range for neutron star binary detection. Comparing with initial detectors, these are called as to `second' generation detectors, including LCGT. LIGO-Australia\cite{LIGOAus} is a project that one set of LIGO instruments will move to western Australia. IndIGo\cite{IndIGo} is an organization of the Indian Initiative in Gravitational-wave Observations.
ET (Einstein Telescope)\cite{ET} is a future plan of gravitational wave detector in Europe, which target more one order better sensitivity than second generation detectors.

\subsection{Detection Range} 

How far astronomical objects can be detected by the gravitational wave detection ? 
The detection range is depend on the gravitational waveform. Here, we discussed about the typical cases: compact binary coalescence and black hole quasi-normal mode oscillation. 

\begin{figure}[htb]
\begin{center}
\includegraphics[width=80mm]{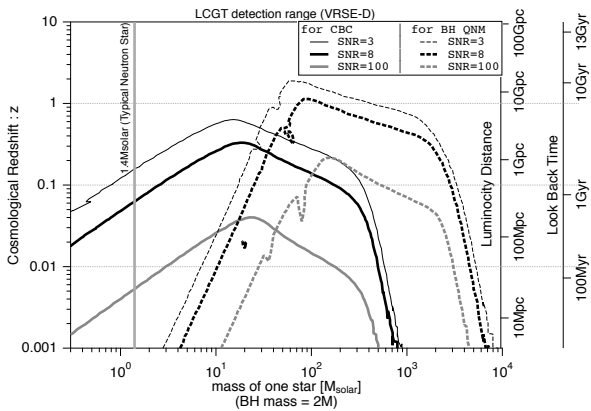}
\end{center}
\vspace{-5mm}
\caption{\label{range}Detection range for LCGT for optimal direction and arrangement of GW source. 
$x$-axis is mass of one star of binary, or half of black hole total mass. Thick solid and thick dashed lines are corresponding to signal-to-noise ratio 8, that is believed as enough signature to claim the detection. 
}
\end{figure}

In case of compact binary coalescence, a gravitational waveform can be predicted with post-newtonian approximation\cite{CBCwaveform}. Using the frequency spectra of gravitational wave and noise power spectral density, the detection range for optimal incident direction and arrangement of compact binaries. Figure \ref{range} displays the detection range as the function of star masses, in case of even mass binary. LCGT's detection range for $1.4 {\rm M_{\odot}}$ binary is about 280 Mpc in optimal case with signal-to-noise ratio 8. Assuming galactic merger rate $118$ events/Galaxy in average and known density of galaxies, LCGT is expected to detect about 10 events per year.

For case of black hole quasi-normal mode, we estimate the signal-to-noise ratio with assumption that 3\% of mass will change as a radiation\cite{FlanaganHughes} and Kerr parameter as 0.9 in this figure.

\section{Global Network of GW detectors}

As we explained previous section, there are some gravitational wave detectors in the world, and these are dislocated. This is important strategy to determine the gravitational wave incident direction and polarization. Global network of the gravitational wave detector is necessary to determine the source direction, to improve whole sky coverage, and extract more information from gravitational wave sources.

Target frequency of the gravitational wave from typical astronomical objects, e.g. compact star coalescence, stellar-core collapse etc. are frequency band of a few 10 Hz to several kHz. Since the wavelength $\lambda$ of 1kHz gravitational wave is 300 km that is longer than visible or infrared light, X or gamma-rays, each ground-based gravitational wave detectors can be treated as a point like. 
We should employ multi-detectors which dislocated longer distance $D$ as 1000 km to determine the direction against the diffraction limit order of $\sim \lambda/D$. The world wide network of gravitational wave detectors can estimate source direction roughly in degree or sub-degree\cite{SRCdirections}.

\begin{figure}
\begin{center}
\includegraphics[width=90mm]{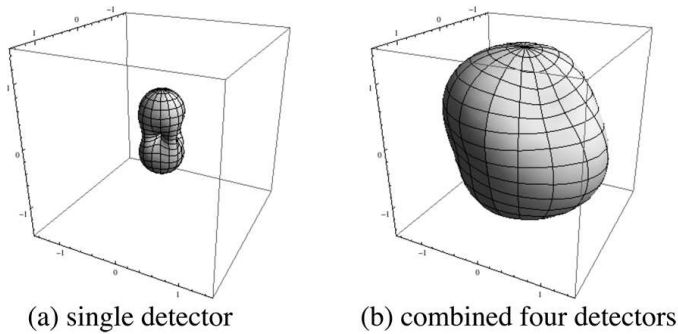}
\end{center}
\vspace{-5mm}
\caption{\label{DetVolume}Antenna pattern of single detector and combined detectors.}
\end{figure}

On the other hand, there is a merit sky coverage with several detectors. A one laser interferometric gravitational wave detector is sensitive for gravitational wave from zenith (and opposite) direction. There is four dead area that corresponding the between interferometer baselines. It is called an antenna pattern of the detector.
Figure \ref{DetVolume} shows (a) single detector's antenna pattern and (b) quadratic sum combined antenna pattern of LCGT, LIGO and Virgo, assuming that all detectors has same sensitivity. Figures are promotional to the survey volume in the universe for non-polarized gravitational wave sources. Average survey range by combined detector is $\sim 2.8$ times of single detector ($22$ times in volume). Quadratic sum of range is very rough estimation, but it displays big advantage in survey volume.
This Combined antenna pattern also remove dead areas with complementary locations of four or more detectors.

\section{Eye and Ear : GW and Counterpart/Follow-up Observations}
Since the gravitational wave is not measured at Summer 2011 yet, we would like to discuss about the prospect of gravitational wave astronomy instead of the summary. 

As we displayed in previous section, a gravitational wave detector's aperture is widely open, almost as a omnidirectional. However, its angular resoletion is poor comparing with optical telescopes. This is a analogy of ``eye and ear''; here, gravitational wave detectors are ears, and optical/radio/X/gamma ray telescopes are eyes. Let's image there is a box that contains something inside. Eyes can see the surface of the box and determine the exact direction of the box, but cannot see the inside. Ears cannot see the box, but suggest what is inside the box by hearing the sound when we shake the box.
Same to this analogy, traditional tools of astronomy and gravitational wave make possible to understand the structure and development of relativistic astronomical objects like as a compact binary, black hole, or supernovae. We hope to open the new window of the astronomy with gravitational waves and counterpart observations.

\section{Acknowledgments}
We would like to thank APRIM organization deeply to invite us the conference.
The author's work was also supported in part by a Monbu Kagakusho Grant-in-aid for Scientific Research of Japan (No. 23540346).


\section*{References}
\begin{thebibliography}{99} 

\bibitem{TaylorHulse}Taylor J. H., et al., Astrophys. J. 345, 434-450

\bibitem{lrr-2008}Duncan R. Lorimer, Living Reviews, lrr-2008-8, http://relativity.livingreviews.org/Articles/lrr-2008-8/

\bibitem{TheLastThreeMin}Cutler C., et al., Phys. Rev. Lett. 70 (1993) 2984

\bibitem{CBCwaveform}Blanchet L., Damour T., Phys.Rev. D46 (1929)4304, \\
Blanchet L., et al., Class. Quantum Grav. 13 (1996) 575

\bibitem{Kim_GMR}Kim, C., AIP Conference Proceedings, vol. 983, pp. 576?583, (American Institute of Physics, Melville, NY, 2008). 

\bibitem{FlanaganHughes} E. E. Flanagan and S. A. Hughes, Phys. Rev. D 57, 4535 (1998)

\bibitem{SRCdirections}L.Wen and Y.Chen, Phys. Rev. D 81, 082001 (2010)

\bibitem{LCGT}K. Kuroda, Classical Quantum Gravity 23, S215 (2006).

\bibitem{LIGO} Rep. Prog. Phys. 72 (2009) 076901

\bibitem{VIRGO} Class. Quantum Grav. 28 114002

\bibitem{LIGOAus}http://www.aigo.org.au/

\bibitem{IndIGo}http://www.gw-indigo.org/

\bibitem{ET}M Punturo et al., Classical and Quantum Gravity, Vol.27 (2010) No.19, 194002

\bibitem{spe_LCGT}http://gwcenter.icrr.u-tokyo.ac.jp/researcher/parameters

\bibitem{spe_aLIGO}https://dcc.ligo.org/cgi-bin/DocDB/ShowDocument?docid=2974

\bibitem{spe_aVirgo}https://wwwcascina.virgo.infn.it/advirgo/

\bibitem{spe_aET}http://www.et-gw.eu/etsensitivities


\end{thebibliography}
\end{document}